\documentclass[10pt,leqno]{amsart}
\usepackage{graphicx}
\usepackage{indentfirst,csquotes}
\usepackage{fancyhdr}
\usepackage[utf8]{inputenc}
\usepackage{natbib}
\usepackage{hyperref}
\usepackage{amsmath,amssymb,amsthm}
\usepackage{titlesec}
\usepackage{lipsum}
\pdfoutput=1

\usepackage{autobreak}
\usepackage{amsfonts}
\usepackage{booktabs}
\usepackage{color}
\usepackage[table]{xcolor}
\usepackage{multirow}
\usepackage{subcaption}
\usepackage[export]{adjustbox}
\usepackage{setspace}

\usepackage{indentfirst,csquotes}
\usepackage{fancyhdr}
\def\bSig\mathbf{\Sigma}

\newcommand{\bfs}{\hbox{\boldmath$s$}}

\newcommand{\bfw}{\hbox{\boldmath$w$}}
\newcommand{\bfz}{\hbox{\boldmath$z$}}

\newcommand{\bftheta}{\hbox{\boldmath$\theta$}}

\usepackage{amssymb,amsthm,amsmath}
\usepackage{xcolor,paralist,hyperref,titlesec,fancyhdr,etoolbox}

\hypersetup{ colorlinks=true, linkcolor=black, filecolor=black, urlcolor=black }

\titleformat{\section}{\normalfont\Large\bfseries}{\thesection}{1em}{}
\titleformat{\subsection}{\normalfont\large\bfseries}{\thesubsection}{1em}{}
\titlespacing*{\section}{0pt}{*3}{*2}
\titlespacing*{\subsection}{0pt}{*2}{*1.5}

\hypersetup{
    colorlinks=true,
    linkcolor=black,
    filecolor=black,
    urlcolor=black
}

\begin{document}
\title{Incorporating Memory into Continuous-Time Spatial Capture-Recapture Models} %%%%%%%%%%%%
\author{\textbf{Clara Panchaud$^{1,*}$, 
Ruth King$^{1}$, David Borchers$^{2,3}$, \\
Hannah Worthington$^{2,3}$, 
Ian Durbach$^{3,4}$, Paul van Dam-Bates$^{2,3,5}$} \\
\\
$^{1}$School of Mathematics and Maxwell Institute, University of Edinburgh, Edinburgh EH9 3FD, Scotland \\
$^{2}$School of Mathematics and Statistics, University of St Andrews, Fife, KY16 9LZ, Scotland \\
$^{3}$Centre for Research into Ecological and Environmental Modelling, \\ University of St Andrews, Fife, KY16 9LZ, Scotland\\
$^{4}$Centre for Statistics in Ecology, the Environment and Conservation, \\ University of Cape Town, Rondebosch 7701, South Africa\\
$^{5}$ Fisheries and Oceans Canada, Pacific Biological Station, Nanaimo, BC, Canada}

\date{\today}

\let\thefootnote\relax
\footnotetext{$^*$Email address: C.E.C.Panchaud@sms.ed.ac.uk}
\begin{abstract}
Obtaining reliable and precise estimates of wildlife species abundance and distribution is essential for the conservation and management of animal populations and natural reserves. Spatial capture-recapture (SCR) models provide estimates of population size and spatial density from data collected from remote sensors such as camera traps. Such data contain spatial correlation between observations of the same individual, which SCR models partly account for through a latent individual-specific activity centre, a location near which the individual is more likely detected. However, SCR models assume that the observations of an individual are independent over time and space, conditional on its activity centre, so that observed sightings at a given time and location do not influence the probability of being seen at future times and/or locations. This assumption is ecologically unrealistic given the smooth movement of animals over space through time. We propose a new continuous-time modelling framework that incorporates both an individual's (latent) activity centre and its (known) previous location and time of detection. By formulating the detections of an individual as an inhomogeneous temporal Poisson process, we develop a model drawing inspiration from the Ornstein-Uhlenbeck process, which is commonly used to model animal movement. Applying our model to a camera-trap survey of American martens, we observe a substantial improvement in model fit and notable differences in the estimated spatial distribution of activity centres. A simulation study shows that standard SCR models can produce substantially biased population estimates when spatio-temporal dependence is ignored, while the memory-based model remains robust. These findings highlight the importance of accounting for memory of previous detections in SCR models to improve ecological interpretation and inference.
\end{abstract}

\maketitle
\newpage

\section{Introduction}
\label{s:introduction}

Determining wildlife population size and distribution within a given area is a crucial goal in ecological research and a key component of effective conservation management. Modern capture-recapture (CR) methods, using data of animals marked when first encountered and then re-observed over time, allow researchers to infer population parameters, often using non-invasive sampling technologies such as camera trapping, acoustic detection, or genetic sampling via hair snares. Data collected from such methods often contain spatial information, making them especially useful for estimating the density of species that are elusive or low in density \citep[e.g.][]{thompson2013sampling}. Among these, camera trap surveys, where individuals are uniquely identifiable from images, have become increasingly popular due to their non-invasive nature, relatively low resource requirements (both financial costs and labour) and ability to operate over long survey periods.

To obtain reliable population size estimates within capture-recapture-type models, it is essential to account for the imperfect detection of individuals in the population \citep{chao2001overview,borchers2002book,gimenez2008risk,king2009book,mccrea2014,king2014,kellner2014accounting,seber2023book}. Spatial capture-recapture (SCR) methods have become a popular framework for inference on population abundance when location data are available. SCR models account for spatial correlation in detections of the same individual by introducing an individual-specific (unobserved) activity centre, a location around which the animal is more likely to be observed \citep{efford2004density,borchers2008spatially}. Extensions of SCR have been developed, including allowing the activity centre locations to vary over time \citep{royle2016spatial}, integrating resource selection information \citep{royle2013integrating}, using continuous detection times rather than discrete survey occasions \citep{borchers2014continuous} and allowing for open animal populations \citep[e.g.][]{gardner2010spatially,glennie2019open}.

Camera trap surveys record the time and location of all detections of an individual. SCR models account for individual spatial heterogeneity through the latent activity centre \citep{borchers2008spatially}. However, they assume that an individual's detections are independent of each other over time, conditional on this activity centre. In other words, the known location of an individual at a previous detection does not inform the likelihood of future detections. This assumption fails to account for the ecological realism that animal movement is continuous and exhibits spatio-temporal correlation, meaning that an animal seen at a trap at a given time is more likely to be observed by nearby traps shortly afterwards, compared to traps at a greater distance. In some cases, SCR estimates can be robust to unmodelled correlation due to animal movement \citep{theng2021confronting}. However, various other sources of heterogeneity in the capture probabilities have previously been shown to lead to biased estimates of population size when ignored in SCR models \citep{moqanaki2021consequences,stevenson2021spatial}, especially when large areas are sampled \citep{howe2022estimating}.

Recent papers have provided methods to consider general spatio-temporal correlations in SCR by including random effects \citep{stevenson2021spatial,dey2022modelling} or by using a hidden Markov model \citep{crum2023forecasting}. \cite{rushing2023ecologist} used a Hawkes process to model clustered observations in time to address non-constant detection rates in a continuous-time non-spatial capture-recapture model, and \cite{kolevextensions} looked into extending the idea to spatial capture-recapture using a Bayesian framework in unpublished work. However, these approaches are computationally expensive to fit and require substantial tuning, limiting their practical use.

In this paper, we propose a new continuous-time spatial capture-recapture approach that introduces spatio-temporal correlation between detections by incorporating a memory component in the detection process. The probability of observing an individual at a given trap is now determined not only by the trap's proximity to the individual's activity centre, as in standard SCR, but also by its proximity to the trap where it was last observed and the time since this observation. We specify this new model within a continuous-time framework, which allows the use of exact observation times and locations to inform the probability of observing an animal in the future, thus making use of the full information from the collected data. We incorporate the memory of past observations through the mean-reverting Ornstein-Uhlenbeck (OU) process, where an individual's activity centre corresponds to the mean of the process. Our model draws inspiration from the OU process as it is commonly used to model animal movement.

We refer to the new proposed model as the memory spatial capture-recapture (MSCR) model. MSCR generalises existing continuous-time SCR models and reduces to the standard formulation as a special limiting case. By incorporating memory, MSCR provides a more ecologically realistic and flexible framework for modelling animal detections in space and time.

The practical relevance of the MSCR model is illustrated using data from a camera trap survey of American martens (\textit{Martes americana}) in New Hampshire, USA. The American marten is an elusive carnivore whose range and abundance have fluctuated over the past century, due to harvesting and habitat loss in the early 1900s, followed by reintroduction programmes with varying success across regions \citep{clark1987,kelly2009,obrien2018}. The species continues to face present and future threats including habitat fragmentation and climate change \citep{hargis1999,carroll2007,lawler2012}. These challenges highlight the need for models that obtain accurate abundance estimates to support effective monitoring and conservation strategies for this species with a complex ecological history.

The survey considered in this paper was conducted over an 11-day period in early 2017 in New Hampshire. Individual martens were identified from their unique chest markings, a method validated by \cite{sirén2016}. The short survey duration make this study particularly well suited to evaluate the benefits of the MSCR framework. Indeed, when repeated detections of the same individual occur over short time intervals, spatio-temporal correlation is more likely to be present and to induce bias in the estimates if unmodelled. In this case study, the MSCR model yielded improved model fit and produced different estimated activity centre locations than the standard SCR model. The results demonstrate the added value of MSCR models to provide inference that can inform the conservation and management of similarly monitored species.

The paper is structured as follows. In Section \ref{s:methods}, we introduce notation and define the MSCR model. The model-fitting procedure is outlined in Section \ref{s:model fitting}. We apply the model to the American martens data set in Section \ref{s:application}, comparing it to the standard SCR model in terms of model fit, parameter estimates and inferred spatial distribution of the individuals. A simulation study is presented in Section \ref{s:simulation}, assessing the performance of MSCR under different levels of spatio-temporal dependence, and comparing it to the standard SCR model. We conclude with a discussion in Section \ref{s:discussion}.

\section{Methods}
\label{s:methods}

\subsection{Notation}

Data are collected at $K$ camera traps arranged in a spatial array within a bounded region $\mathcal{R} \subset \mathbb{R}^2$ of area $A$, over a fixed time interval $[0,T]$. The camera traps are assumed to be continuously active throughout the study period, but the approach can be immediately generalised to include temporary deactivation periods. The locations of the $K$ camera traps are denoted by $\{\bfz_k \in \mathcal{R}; k=1,\dots,K\}$. We assume that the visibility regions of the traps are non-overlapping, so that an animal can be observed by at most one camera at any given instant. We let $N$ denote the (unknown) total population size of individuals within region $\mathcal{R}$. The population is assumed closed during the interval $[0,T]$. The primary aim is to estimate this total population size $N$.

The data consist of the capture histories for $n \le N$ observed individuals. Each observed individual's capture history is a set of (continuous) times and camera locations at which an individual was detected. Note that we use the terms capture, observation and detection interchangeably. We assume that each individual is uniquely and correctly identified at each capture, allowing individual-specific capture histories to be constructed. Each observed individual $i$ has a latent activity centre $\bfs_i \in \mathcal{R}$. Let $J_i$ denote the number of times individual $i$ is observed. The $j^{th}$ observation of individual $i$ is represented as a vector $\bfw_{ij} = (t_{ij}, k_{ij})$, where $t_{ij}$ is the detection time and $k_{ij}$ the index of the trap that recorded the observation. The full capture history of individual $i$ is then $\bfw_i=(\bfw_{i1},...,\bfw_{iJ_i})$. The observed data consist of all $n$ observed capture histories, denoted by $\bfw = (\bfw_1,\dots,\bfw_n)$.

\subsection{Model formulation}
\label{s:likelihood_formulation}

The primary parameter of interest is the total population size, $N$. Let $\bftheta$ denote the remaining model parameters needing to be estimated. We adopt a likelihood-based approach, following the formulation from \cite{borchers2014continuous} which leads to
\begin{equation*}
\label{General_likelihood}
    L(\bftheta, N ; n,\bfw)\propto f(n ; \bftheta, N)f(\bfw ; n,\bftheta, N ),
\end{equation*}
where we assume $n$ is a binomial random variable and capture histories between individuals are independent such that 
\begin{equation}
\label{data_lik}
 f(\bfw ; n,\bftheta, N ) = \prod_{i=1}^n \frac{f(\bfw_i;\bftheta)}{p(\bftheta)},
\end{equation}
where $p(\bftheta)$ is the probability than an individual in the population is observed at least once during the survey. We consider a survival analysis formulation, in which the detections of a given individual are generated by an inhomogeneous temporal Poisson process. Under this framework, an event corresponds to an observation so that "surviving" over a time period means that an individual is not detected by any trap during that period. Each individual's likelihood contribution is defined in the traditional survival analysis framework and integrated over the potential activity centre locations. Detailed derivations are provided in Appendix \ref{AppendixA}. Survival analysis is based on hazard functions which we will now introduce.

\subsection{Hazard function}
\label{s:hazard}

The hazard function corresponds to the instantaneous rate of a detection for an individual by a given trap at a specified time. We define the hazard function in terms of both the (unknown) activity centre for the individual and its most recent known location and time (if previously observed by a camera trap), thus incorporating known sighting information. In particular, we specify the hazard function through the distribution of an Ornstein-Uhlenbeck (OU) process \citep{uhlenbeck1930theory}, commonly used to model animal movement as it provides a good approximation of smooth movement through time \citep{dunn1977analysis,blackwell1997diffusion}.  The hazard function for location $\bfz \in \mathcal{R}$ at time $t$, given an activity centre, $\bfs \in \mathcal{R}$, and the last previous observed location $\bfz^*$ that happened at time $t^*<t$ is given by,
\begin{equation}
    \label{OU_hazard}
    \begin{aligned}
        &h(\bfz,t;\bfs,\bfz^*,t^*,\bftheta) =\\ &h_0\exp \left[ -\frac{1}{2} \Bigl\{\bfz-\boldsymbol{\mu}(t; \bfs, \bfz^*, t^*, \beta)\Bigl\}^{\top}\Sigma(t; t^*,\sigma^2,\beta)^{-1} \Bigl\{\bfz-\boldsymbol{\mu}(t; \bfs, \bfz^*,t^*, \beta)\Bigl\}\right],
    \end{aligned}
\end{equation}
where $\bftheta = \{h_0,\sigma^2,\beta\}$ denotes the set of parameters to be estimated. The mean and covariance matrices are defined such that,
$$\boldsymbol{\mu}(t; \bfs, \bfz^*,t^*, \beta)= e^{-\beta (t-t^*)}\bfz^*+ \bigl\{1-e^{-\beta (t-t^*)}\bigl\}\bfs,$$
and
$$\Sigma(t;t^*,\sigma^2,\beta) = \sigma^2 \bigl\{1 - e^{-2\beta (t-t^*)} \bigl\}I_2.$$
It is only necessary to evaluate the hazard function at the $K$ locations of the traps, because the observed locations of individuals will always correspond to trap locations (as individuals are only recorded at camera trap locations), so that $\bfz$ always corresponds to one of the camera trap locations, i.e. $\bfz\in{\bfz_k; k=1,\dots,K}$.

The model parameters consist of an intercept parameter, $h_0$, a scale parameter, $\sigma^2$, and a weight parameter, $\beta>0$, that determines the balance between attraction to the activity centre and persistence within an area. Small values of $\beta$ correspond to a weaker (or slower) attraction of individuals to the area closer to their activity centre; while larger values of $\beta$ correspond to a stronger (or faster) attraction towards their activity centre. As $\beta \rightarrow \infty$ the hazard function from Equation \ref{OU_hazard} reduces to a standard half-normal form,
\begin{equation}
    \label{halfnormal}
    \begin{aligned}
        \lim_{\beta\rightarrow \infty} h(\bfz, t;\bfs,\bfz^*,t^*,\bftheta) &\equiv 
        hn(\bfz;\bfs,h_0,\sigma^2) &= \displaystyle h_0  \exp \Bigl\{-\frac{ (\bfz - \bfs)^{\top} (\bfz - \bfs)}{2\sigma^2}\Bigl\},
    \end{aligned}
\end{equation}
and does not depend on any previous time or location (or current time). The half-normal hazard function, $hn$, is the hazard function used by \cite{distiller2020using} in the standard continuous-time SCR model. We refer to the standard continuous-time SCR model as CT SCR from here onwards for simplicity. When referring to CT SCR, the parameters to be estimated are $\bftheta=(h_0,\sigma^2)$ as in Equation \ref{halfnormal}. As the forms of the hazard functions of MSCR and CT SCR differ, the parameters $h_0$ and $\sigma^2$ of each model have different interpretations. Equation \ref{halfnormal} shows that CT SCR is a limiting case of our MSCR model as $\beta\rightarrow \infty$.

We also recover the half-normal hazard function when the time difference $(t-t^*) \rightarrow \infty$. Consequently, the previous known location of an individual at time $t^*$ does not influence its current rate of detection if it corresponds to a location that the animal visited a very long time ago.

The camera traps are regarded as switched off until the start of the survey. As a result, the hazard function of an individual that has not yet been observed is equal to its limiting distribution, $hn(\bfz;\bfs,h_0,\sigma^2)$, as in Equation \ref{halfnormal}. Equivalently, we can consider an individual that has not been seen yet as if last observed at $t=-\infty$.

\section{Model fitting} 
\label{s:model fitting}

To fit the MSCR model to data and estimate abundance, we adopt a conditional likelihood approach \citep{borchers2002book,borchers2008spatially}. We discuss the two stages of this approach in turn.

\subsection*{Step 1: Estimation of $\widehat{\bftheta}$} 

The parameters $\widehat{\bftheta}$ are estimated by maximising the likelihood term given in Equation (\ref{data_lik}). However, this expression is analytically intractable due to the integrals over both time (appearing in the survival function) and space (in relation to the latent activity centres). These integrals are approximated numerically using a piecewise-constant discretisation with a midpoint rule. Time is partitioned into equal-length intervals between observations, while space is discretised using a landscape mask, a common approach in SCR studies.

\subsection*{Step 2: Estimation of $\widehat{N}$} 

Given the parameter estimate $\widehat{\bftheta}$, the total population size, $\widehat{N}$, is estimated using the Horvitz–Thompson-type estimator,

\begin{equation*}
    \widehat{N}=\frac{n}{p(\widehat{\bftheta})},
\end{equation*}
where we recall that $n$ denotes the number of observed individuals within the study; and $p(\widehat{\bftheta})$ denotes the probability an individual is observed at least once during the survey, evaluated at the MLE of the parameters, $\widehat{\bftheta}$. The variance of $\widehat{N}$ is obtained following the approach of \cite{huggins1989statistical} and \cite{alho1990logistic}, as 
\begin{equation*}
    \hbox{Var}(\widehat{N}(\widehat{\bftheta}))= \left[  \frac{ \partial \widehat{N}(\widehat{\bftheta})  }{ \partial\widehat{\bftheta}  } \right]  \hbox{Cov}(\widehat{\bftheta})  \left[  \frac{ \partial \widehat{N} (\widehat{\bftheta})  }{ \partial\widehat{\bftheta}} \right] ^{\top} + \frac{n (1-p(\widehat{\bftheta}))}{p(\widehat{\bftheta})^2}.
\end{equation*}

If inference on the individual activity centre locations is of interest, the posterior density for the location $\bfs_i$ of individual $i$ can be obtained through Bayes' theorem, following \cite{borchers2008spatially}, as"
\begin{equation}
    \label{e:ac_pdf}
    f(\bfs_i; \bfw_i,\bftheta) \propto  \frac{ f(\bfw_i ;\bfs_i,\bftheta)}{ \int_{\mathcal{R}} f(\bfw_i ; \bfs, \bftheta) d\bfs},
\end{equation}
with the MLE $\widehat{\bfs_i}$ defined as the location with the highest density. 

Model fitting was implemented in \texttt{R} \citep{R}. All analyses were conducted in RStudio on a MacBook Air (M1, 2020) with 16 GB of memory under macOS Sonoma. The code is made available for reproducibility in the supplementary materials.

\section{Application: American Martens}
\label{s:application}

\begin{figure}
    \centering
    \includegraphics[width=6in]{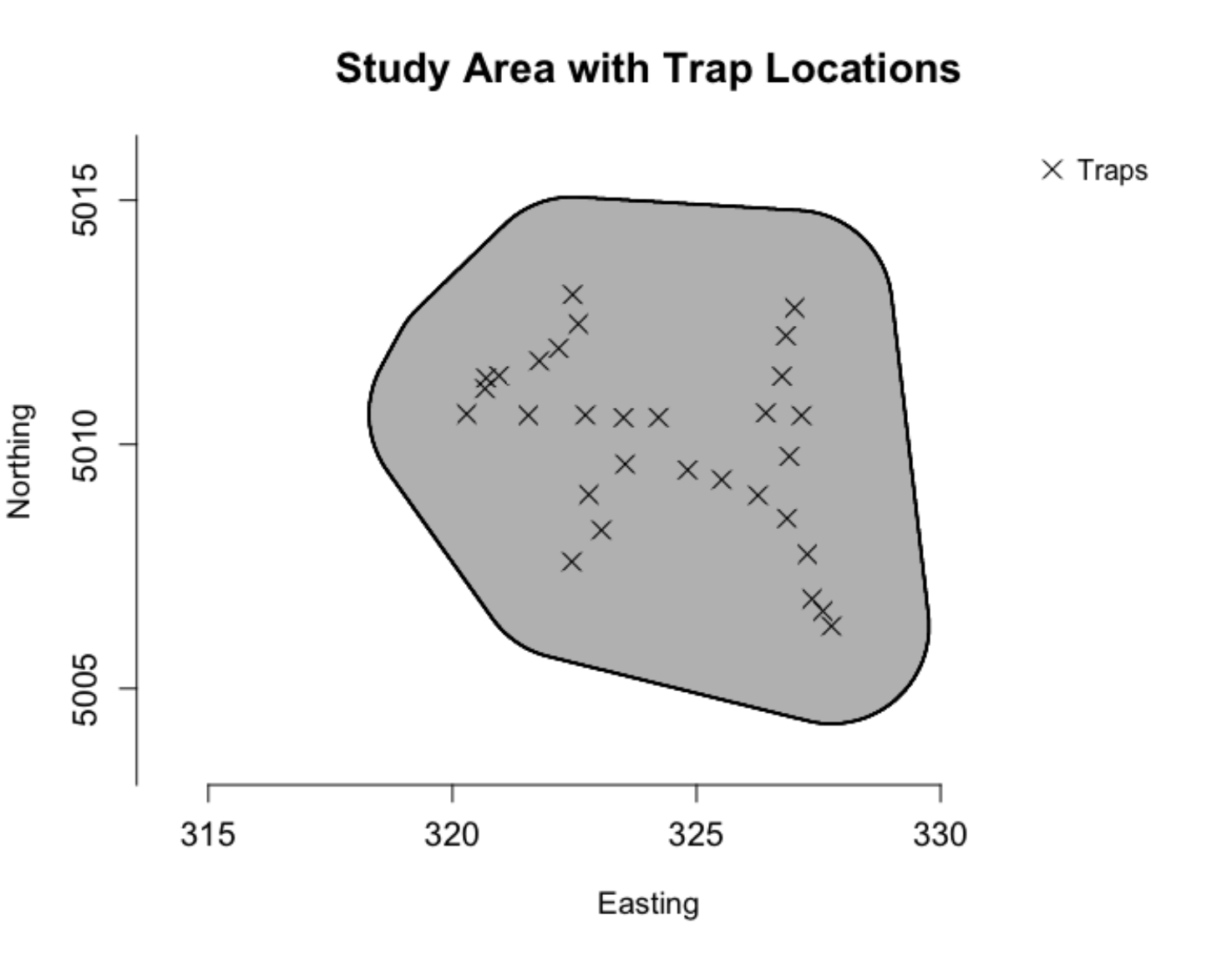}
    \caption{Study area of the American marten survey. Map units are in kilometres. Camera trap locations are indicated by the crosses and the region is defined with a 2-km buffer around the trap array.}
    \label{f:landscape}
\end{figure}

We analyse data collected from a camera trap survey conducted in 2017 to monitor American martens (\textit{Martes americana}) at a study site in northern New Hampshire, along the United States–Canada border. The survey deployed 30 camera traps to survey the study area for a period of 11 days, between the 28th of March and the 7th of April 2017. The short duration of the survey increases the likelihood of spatio-temporal correlation in repeated detections of the same individual, motivating the use of the MSCR model. The survey detected a total of 9 individuals, observed on average over 8 times, 5 of whom were detected by multiple cameras. The study area, illustrated in Figure \ref{f:landscape}, includes a 2 km buffer around the convex hull of the traps, resulting in a total area of 95.44 km$^2$.

We fit both the MSCR model and the CT SCR model to the data. The numerical integration over space was performed using a mesh with 2386 grid points. For the integration over time we used discrete intervals of 2.64 hours duration. Alternative discretisation schemes were explored to ensure numerical stability and computational efficiency. Fitting the models to the data took 9.86 and 4.37 minutes for MSCR and CT SCR respectively.

\begin{table}
  \centering
  \begin{subtable}[t]{\linewidth}
    \centering
    \begin{tabular}{ccc}
      \toprule
      \multicolumn{3}{c}{\textbf{MSCR Model}} \\
      \multicolumn{3}{c}{$\Delta$AIC = 0} \\
      \midrule
      Parameter & Estimate (SE) & 95\% CI \\ 
      \midrule
      $N$ & 12.57 (2.50) & (8.55, 18.50) \\ 
      $h_0$ & 2.27 (0.48) & (1.50, 3.45) \\
      $\sigma^2$ & 0.32 (0.063) & (0.22, 0.47) \\
      $\beta$ & 2.07 (0.45) & (1.35, 3.16) \\ 
      \bottomrule
    \end{tabular}

    \caption{Parameter estimates from fitting the MSCR model to the American marten data set.}
    \label{t:American1}
  \end{subtable}
  
  \vspace{1em}
  
  \begin{subtable}[t]{\linewidth}
    \centering
    \begin{tabular}{ccc}
      \toprule
      \multicolumn{3}{c}{\textbf{CT SCR Model}} \\
      \multicolumn{3}{c}{$\Delta$AIC = 97.98} \\
      \midrule
      Parameter & Estimate (SE) & 95\% CI \\ 
      \midrule
      $N$ & 14.40 (3.09) & (9.50, 21.83) \\ 
      $h_0$ & 1.60 (0.60) & (0.77, 3.32) \\
      $\sigma^2$ & 0.27 (0.050) & (0.18, 0.39) \\
      \bottomrule
    \end{tabular}
    \caption{Parameter estimates from fitting the CT SCR model to the American marten data set.}
    \label{t:American2}
  \end{subtable}

  \caption{Parameter estimates, standard errors (SE) and 95\% confidence intervals (CI) from fitting the MSCR and CT SCR models to the American marten data set.}
  \label{t:Americanmartens}  
\end{table}

Table \ref{t:Americanmartens} provides the estimates of the model parameters and population size for both the MSCR and CT SCR models. The AIC statistic \citep{akaike1973} for the MSCR model is substantially lower than for the CT SCR model ($\Delta$AIC$=97.98$), indicating a superior fit of the model to the data with the inclusion of the memory component. The corresponding estimates of the total population size are relatively similar across the two different models, with an estimate of 12.57 for MSCR and 14.40 for CT SCR. While relatively close, the CT SCR estimate may be biased upward due to unmodelled spatio-temporal correlation. The 95\% confidence interval (CI) for MSCR (width = 9.95) is narrower than that of CT SCR (width = 12.33), despite the MSCR model having one additional parameter.

\begin{figure}[htbp!]
  \captionsetup[subfigure]{aboveskip=0.1pt, belowskip=0.1pt}

  \centering

  \begin{subfigure}[t]{0.98\linewidth}
    \includegraphics[width=\linewidth]{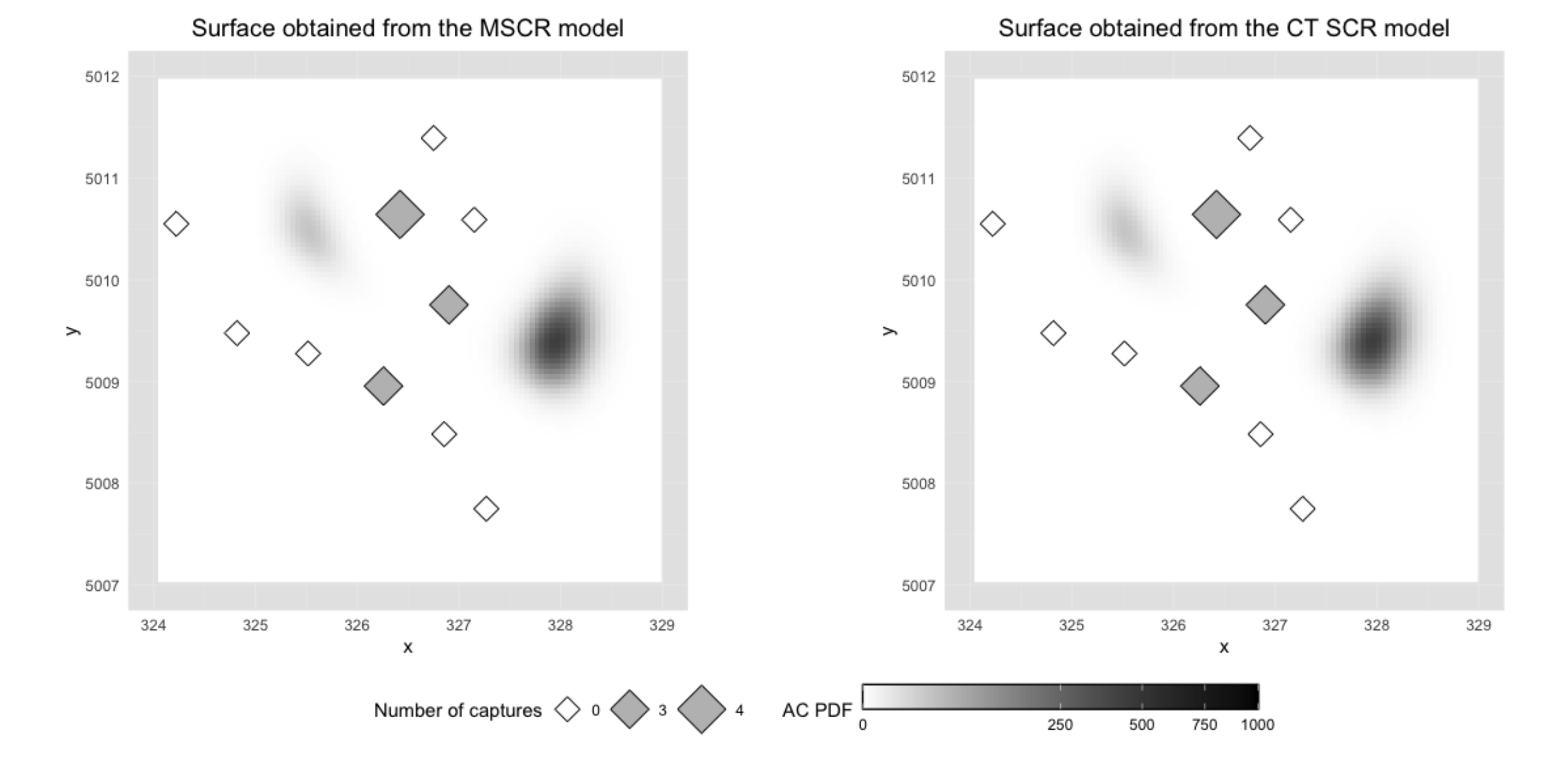}
    \caption{Estimated surface obtained from both models for an individual American marten.}
    \label{f:surface1}
  \end{subfigure}

  \begin{subfigure}[t]{0.98\linewidth}
    \includegraphics[width=\linewidth]{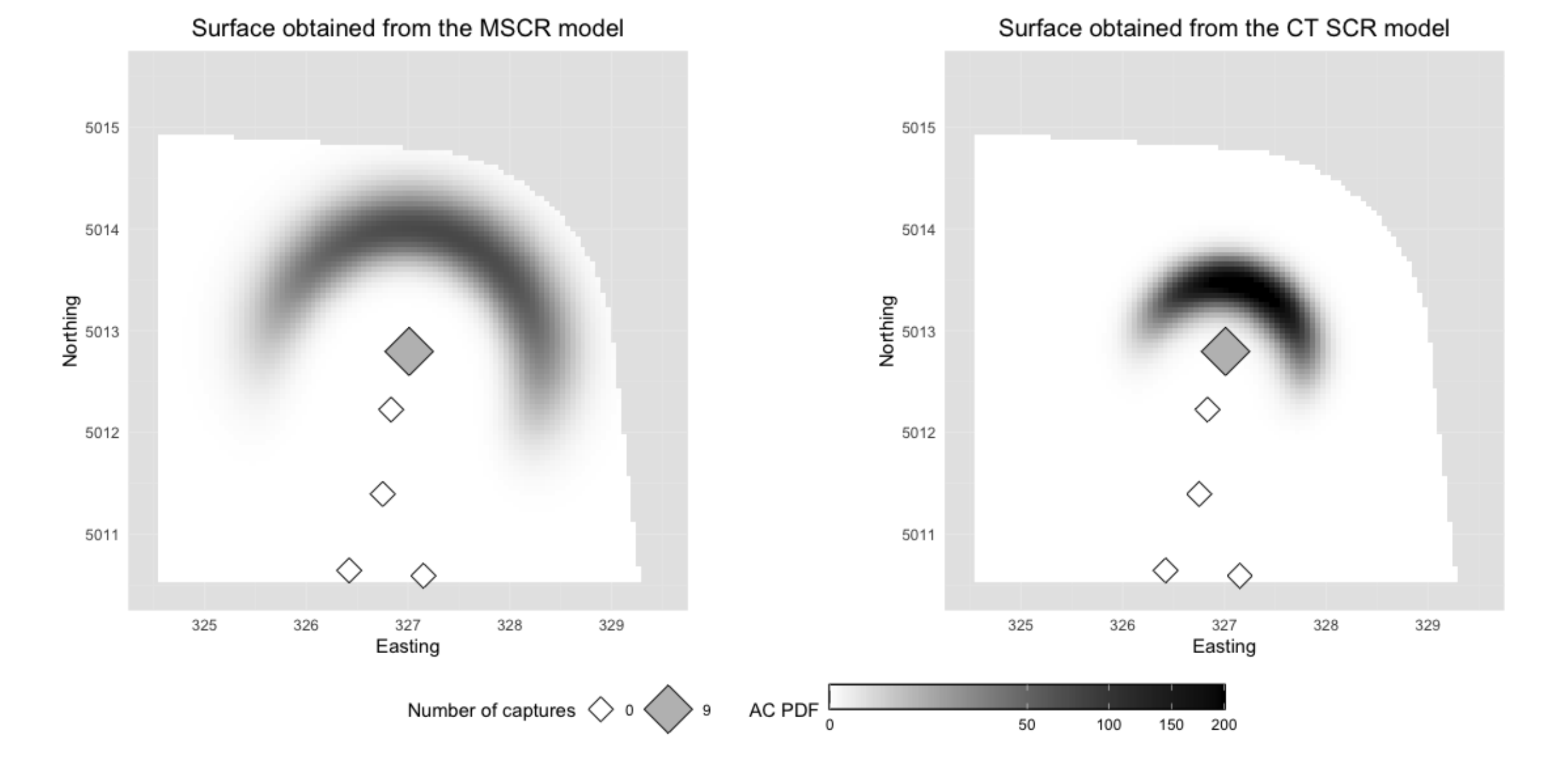}
    \caption{Estimated surface obtained from both models for a second individual American marten.}
    \label{f:surface2}
  \end{subfigure}

  \begin{subfigure}[t]{0.98\linewidth}
    \includegraphics[width=\linewidth]{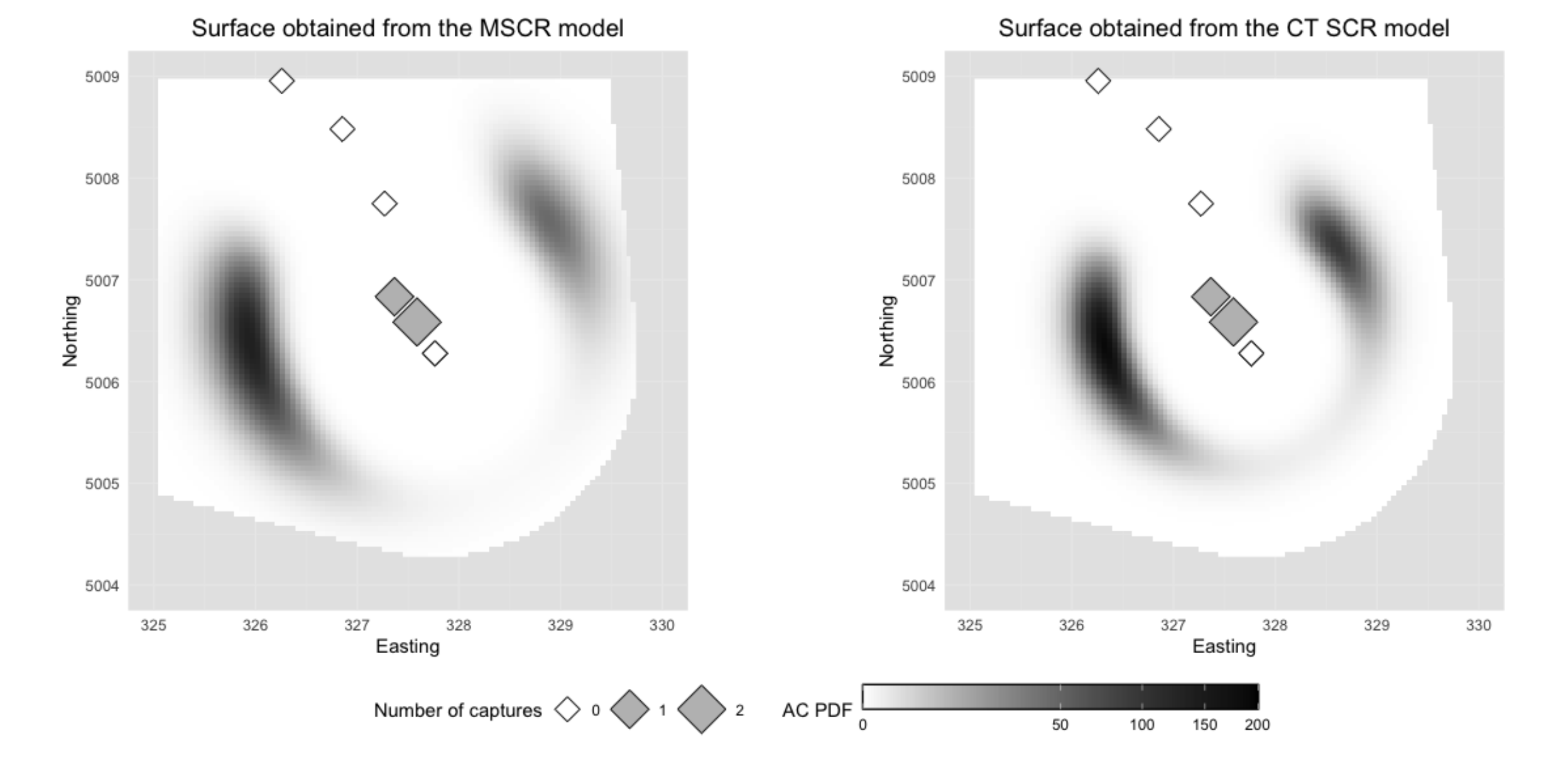}
    \caption{Estimated surface obtained from both models for a third individual American marten.}
    \label{f:surface3}
  \end{subfigure}

  \caption{Estimated activity centre probability density surfaces from the MSCR and CT SCR models for three example individuals in the American marten data set. “AC PDF” denotes the activity centre probability density function. Map units are in kilometres.}
 \label{f:ac_plots}
\end{figure}

Figure \ref{f:ac_plots} displays the estimated activity centre probability density surfaces for three observed individuals. Substantial differences are observed between the MSCR and CT SCR models, with minimal overlap in regions of high-density. In Figure \ref{f:surface1}, for example, the MSCR and CT SCR models concentrate density in entirely different areas. Figure \ref{f:surface2} as well as \ref{f:surface3}, highlight that the MSCR model assigns higher density farther from recent trap detections, while CT SCR concentrates density near traps with repeated detections in a short time span. By accounting for spatio-temporal correlation, MSCR allows for the possibility that repeated detections at a given trap reflect short-term movement through the local area, rather than proximity to the animal’s activity centre. The differences in estimated activity centre locations may have important implications for ecological inference and conservation decisions, as they indicate where the animals live. The differences are likely to be more pronounced, and therefore more impactful, in short-duration surveys where detections tend to exhibit more spatio-temporal correlation.

\section{Simulation Study} 
\label{s:simulation}

To evaluate the benefit of incorporating memory into spatial capture-recapture, we conduct a simulation study comparing three models: our proposed memory-based SCR model (MSCR), the continuous-time SCR model (CT SCR), and the standard discrete-time SCR model (DT SCR). Data are generated through the MSCR model using the trap array and region $\mathcal{R}$ introduced in Section \ref{s:application}, based on the American marten survey. We include the 2 km buffer around the traps, resulting in the region from Figure \ref{f:landscape}. The simulated survey duration is $T=11$ days, again as in the American martens case. The true populations size is fixed at $N=20$ to produce data sets with slightly larger sample sizes than in the American marten study. The population size is treated as a fixed value under a binomial observation model.

We set $h_0 = 2.27$ and $\sigma^2 = 0.32$ to match the parameter estimates from the case study, ensuring that the simulated data reflect realistic conditions. The parameter $\beta$ controls the strength of the individual's tendency to remain near their previous location; higher value correspond to a faster return to the area surrounding their activity centre, while lower values reflect stronger site fidelity. To explore a range of memory effects and their impact on the results of the different models, we vary $\log(\beta)$ across the values $(-1.5, -1, -0.5, 0, 0.5, 1, 1.5)$. We simulate 100 data sets for each value of $\beta$.

Each simulated individual is assigned an activity centre uniformly drawn from the region $\mathcal{R}$ and a starting location drawn from a half-normal distribution around that activity centre. To simulate a capture history we discretize the survey time from 0 to $T$ in fine intervals of length $T/1000$, assuming a piecewise constant hazard function within each interval and use the associated all-traps hazard function to determine if a capture occurred in each interval. If there is a capture in the interval, the capture time is set to be the associated end-point of the given interval, and the corresponding trap location drawn such that the probability associated with each trap location is proportional to their relative trap hazard functions. The individual's hazard function is updated at each interval, based on the elapsed time and the most recent observed location. On average, $71\%$ of individuals (approximately 14 individuals per data set) were detected at least once.

We fit the MSCR model to each simulated data set using a mesh of 381 grid points and a time discretisation with intervals of length $T/20$. Given the large number of simulations (700 in total), we adopted a coarser discretisation of space and time than in the case study to reduce computational effort. This discretisation was selected to ensure the accuracy of the estimates while maintaining computational feasibility. We compare the results of MSCR with

\begin{itemize}
    \item CT SCR, the continuous-time SCR model without memory (the limiting case of MSCR as $\beta \rightarrow \infty$). Following common practice, we thin this data to at most one observation per hour.
    \item DT SCR, the discrete-time SCR model, commonly used in practice to analyze such data. We collapse the data into counts over daily capture occasions. DT SCR is fitted using the R package \texttt{secr} \citep{secr}.
\end{itemize}

\begin{figure}[htbp!]
    \centering
        \includegraphics[width=\linewidth]{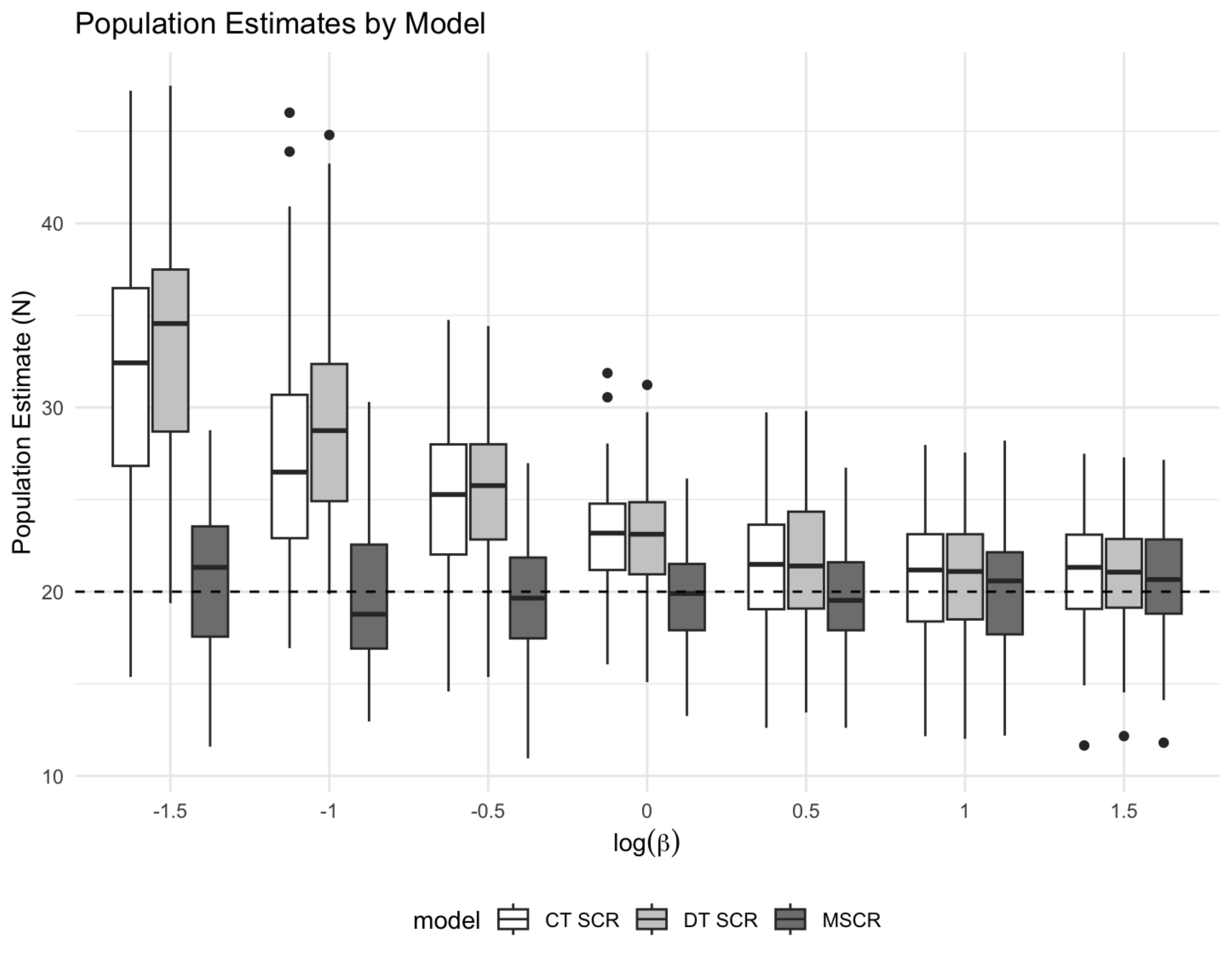}
        \caption{ Population size estimates ($\widehat{N}$) across 100 simulated data sets for each value of $\beta$. Estimates are shown for the MSCR, CT SCR, and DT SCR models. The dashed line indicates the true population size ($N=20$).}
    \label{f:boxplot}
\end{figure}

Table \ref{t:simulations} summarizes the performance of all three models across varying memory strengths. It reports the average population size estimates, bias, uncertainty, coverage and RMSE. Boxplots of the population size estimates for each model are displayed in Figure \ref{f:boxplot}. The average model fitting times were 187 seconds for MSCR; 55 seconds for CT SCR and 0.10 seconds for DT SCR using a MacBook Air (M1, 2020) with 16 GB of memory under macOS Sonoma.

\begin{table}
\hspace{-1cm}
\centering
\setlength{\tabcolsep}{3.5pt}  
\small  
\resizebox{\linewidth}{!}{
\begin{tabular}{cccccccccc}
\toprule
Model & Metric & \multicolumn{7}{c}{$\log(\beta)$ values} \\
\cmidrule(lr){3-9}
 &  & -1.5 & -1 & -0.5 & 0 & 0.5 & 1 & 1.5 \\
\midrule
\multirow{5}{*}{MSCR} 
 & Estimate (SE) & 20.75 (4.17) & 19.67 (3.41) & 19.52 (3.09) & 19.71 (3.02) & 19.76 (2.95) & 20.13 (3.03) & 20.56 (2.99) \\
 & \% Bias & 3.75 & -1.65 & -2.40 & -1.45 & -1.20 & 0.65 & 2.80 \\
 & 95\% CI Width & 16.36 & 13.38 & 12.13 & 11.84 & 11.57 & 11.90 & 11.71 \\
 & \% Coverage & 90 & 90 & 88 & 92 & 94 & 95 & 93 \\
 & RMSE & 4.38 & 3.67 & 3.21 & 2.91 & 3.01 & 3.16 & 3.12 \\
\midrule
\multirow{5}{*}{CT SCR} 
 & Estimate (SE) & 31.81 (6.58) & 27.08 (5.05) & 24.81 (4.33) & 22.83 (3.79) & 21.26 (3.29) & 20.76 (3.18) & 21.02 (3.12) \\
 & \% Bias & 59.05 & 35.40 & 24.05 & 14.15 & 6.30 & 3.80 & 5.10 \\
 & 95\% CI Width & 25.80 & 19.81 & 16.98 & 14.86 & 12.91 & 12.45 & 12.21 \\
 & \% Coverage & 55 & 75 & 90 & 98 & 95 & 92 & 92 \\
 & RMSE & 13.70 & 9.22 & 6.31 & 4.30 & 3.53 & 3.42 & 3.33 \\
\midrule
\multirow{5}{*}{DT SCR} 
 & Estimate (SE) & 33.41 (9.12) & 28.88 (7.76) & 25.43 (6.77) & 22.86 (6.10) & 21.47 (5.71) & 20.74 (5.54) & 20.93 (5.51) \\
 & \% Bias & 67.05 & 44.40 & 27.15 & 14.30 & 7.35 & 3.70 & 4.65 \\
 & 95\% CI Width & 35.74 & 30.40 & 26.52 & 23.90 & 22.39 & 21.71 & 21.58 \\
 & \% Coverage & 86 & 95 & 100 & 100 & 100 & 100 & 100 \\
 & RMSE & 14.84 & 10.27 & 6.73 & 4.25 & 3.74 & 3.43 & 3.17 \\
\bottomrule
\end{tabular}
}

\caption{Population size estimates ($\widehat{N}$) across simulation studies with different values of $\log(\beta)$. The average estimated $\widehat{N}$ over the 100 data sets first displayed, along with the average estimated standard error, followed by \textit{\% Bias} giving the average bias of the estimates, \textit{95\% CI width} and \textit{\% Coverage} providing information on the estimated confidence intervals and \textit{RMSE} being the root mean square error.
}
\label{t:simulations}
\end{table}

From Table \ref{t:simulations} we see that the MSCR model provides accurate population size estimates for all values of $\beta$ with a bias consistently under $4\%$. In contrast, the continuous and discrete SCR both present a large positive bias for low values of $\beta$, such as a bias of $59.05\%$ from CT SCR and $67.05\%$ for DT SCR when $\log(\beta)=-1.5$. As the value of $\beta$ increases, the estimates for MSCR and CT SCR become increasingly similar, as expected, and the bias of CT SCR reduces although for all values of $\beta$ presented here it remains larger than the bias of MSCR. The parameter estimates of CT SCR and DT SCR are highly similar for all $\beta$s, consistent with the results of \cite{distiller2020using}.

Furthermore, the RMSE values for MSCR remain consistently low, indicating strong model performance across different values of $\beta$. However, RMSE is significantly higher for CT and DT SCR at lower $\beta$ values. For $\log(\beta)=-1.5$, omitting the memory component and fitting the CT SCR model reduces the coverage probability to $55\%$, compared to $90\%$ coverage of MSCR. As $\beta$ increases, the coverage of CT SCR stabilizes and becomes more accurate. On the other hand, DT SCR tends to provide an over-coverage of the population size for most values of $\beta$, with wider confidence intervals. Overall, MSCR, with its additional memory component, substantially improves population size estimation and reduces bias. The RMSE of $N$ for MSCR is approximately one-third that of either SCR model. Figure \ref{f:boxplot} displays the estimated population sizes across the 100 simulated data sets for each model and $\beta$ value. Both SCR models increasingly overestimate the population size as $log(\beta)$ decreases.

\begin{figure*}[htbp!]
    \centering
    \captionsetup{skip=2pt}

    \begin{subfigure}[t]{0.495\linewidth}
        \includegraphics[width=\linewidth]{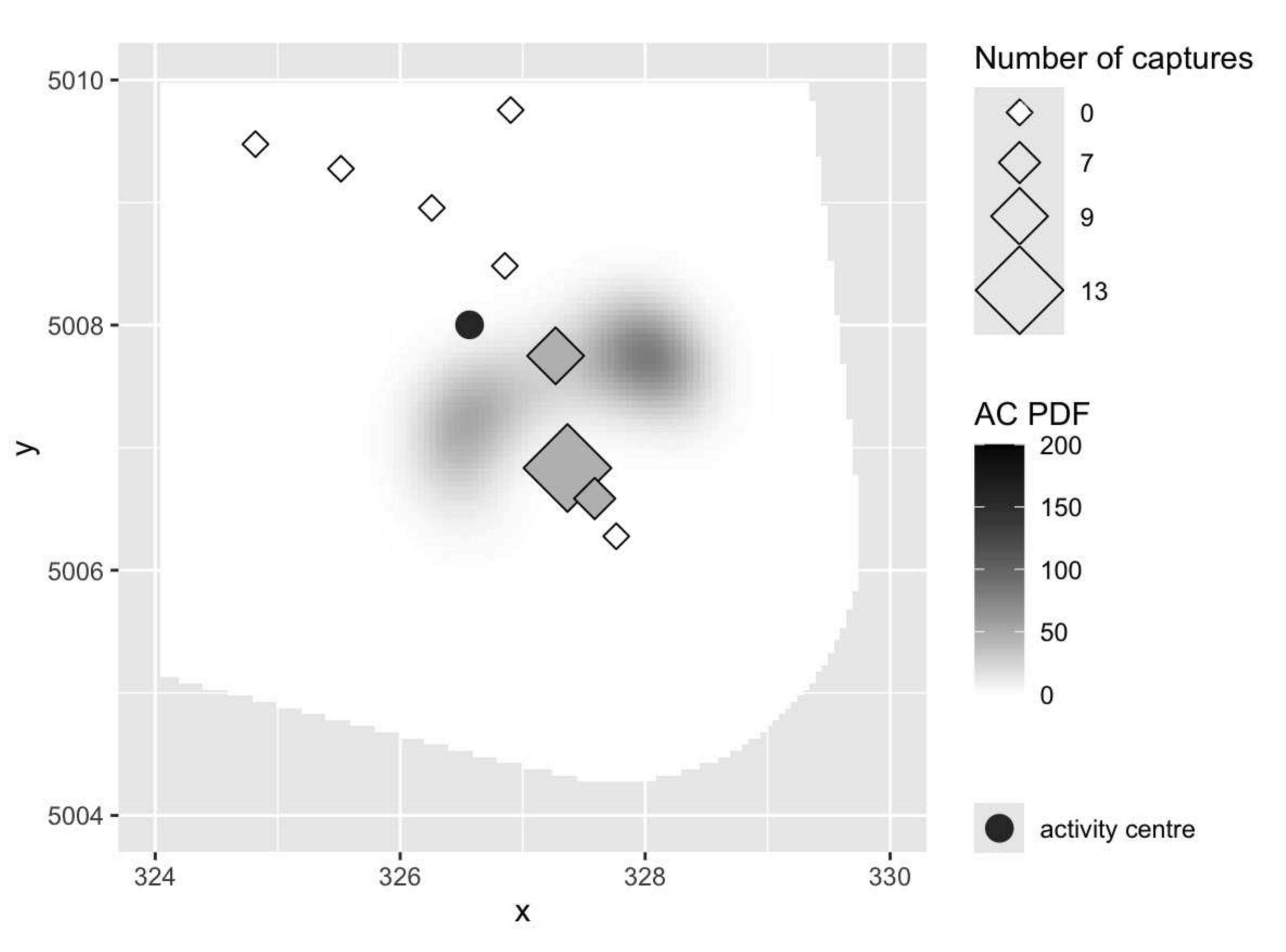}
        \caption{Estimated surface obtained from the MSCR model for a simulated individual.}
        \label{f:sim_surface1}
    \end{subfigure}%
    \hfill
    \begin{subfigure}[t]{0.495\linewidth}
        \includegraphics[width=\linewidth]{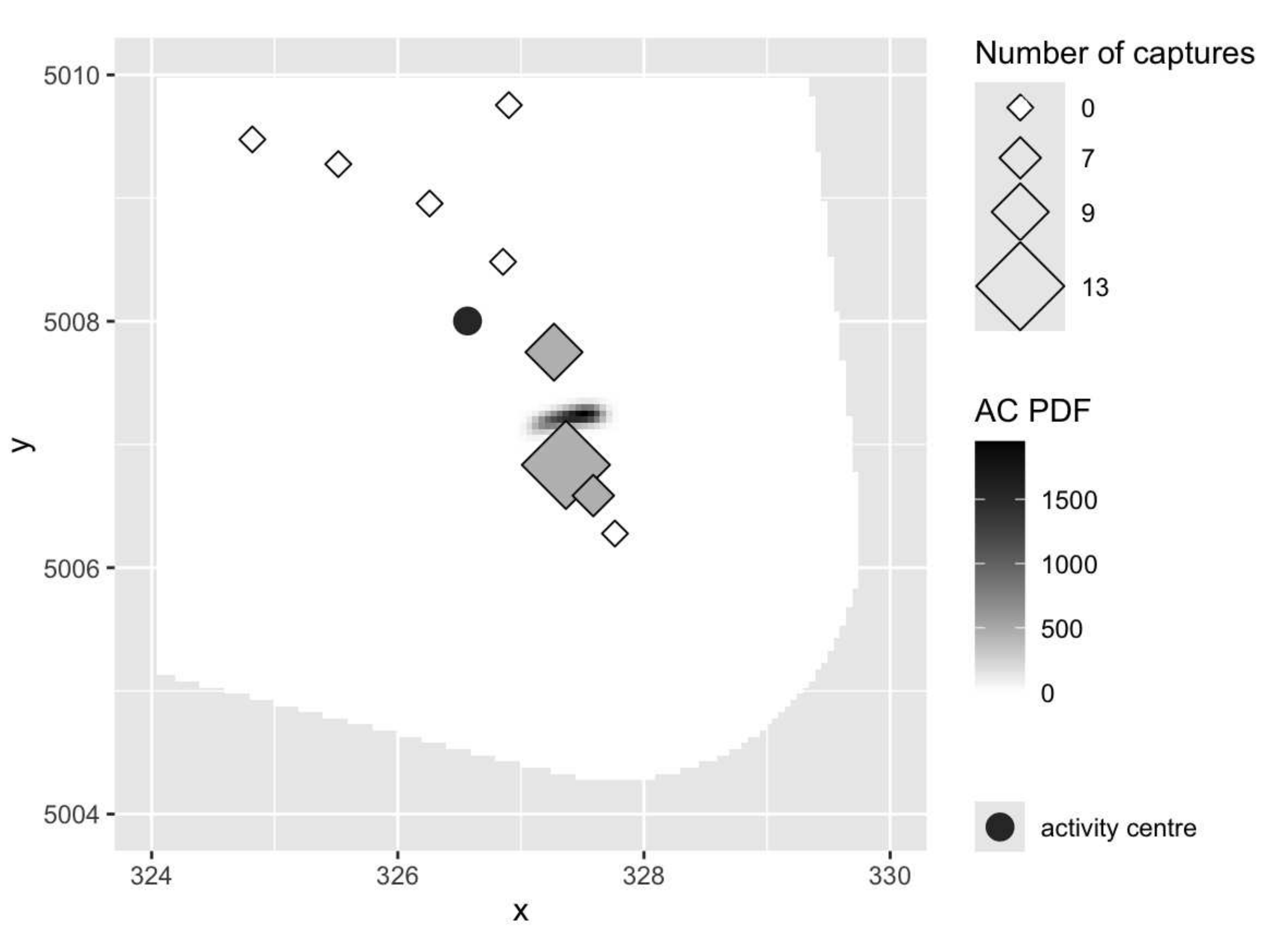}
        \caption{Estimated surface obtained from the CT SCR model for the same simulated individual as in (a).}
        \label{f:sim_surface1_nm}
    \end{subfigure}

    \vspace{10pt}

    \begin{subfigure}[t]{0.495\linewidth}
        \includegraphics[width=\linewidth]{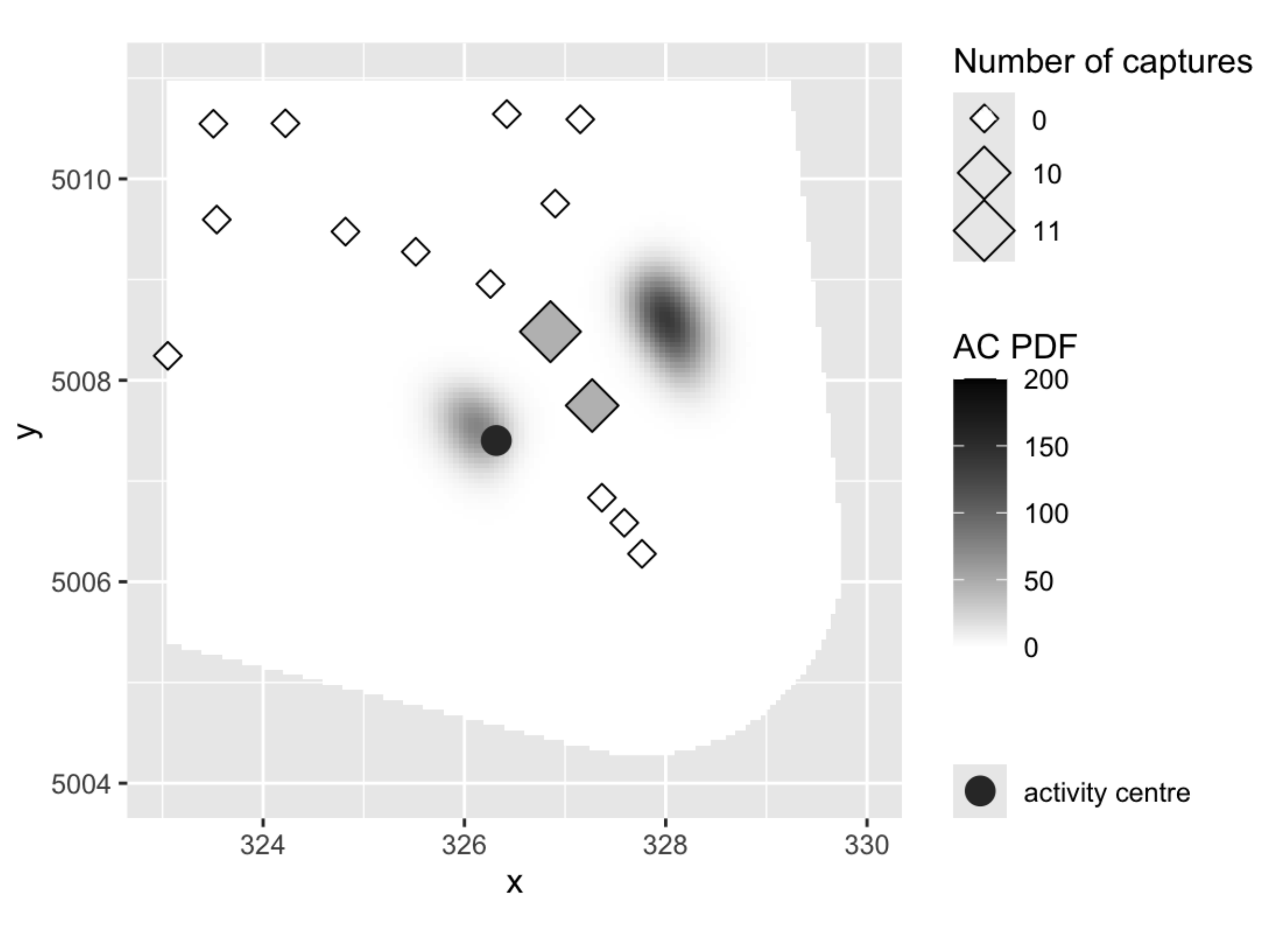}
        \caption{Estimated surface obtained from the MSCR model for a simulated individual.}
        \label{f:sim_surface2}
    \end{subfigure}%
    \hfill
    \begin{subfigure}[t]{0.495\linewidth}
        \includegraphics[width=\linewidth]{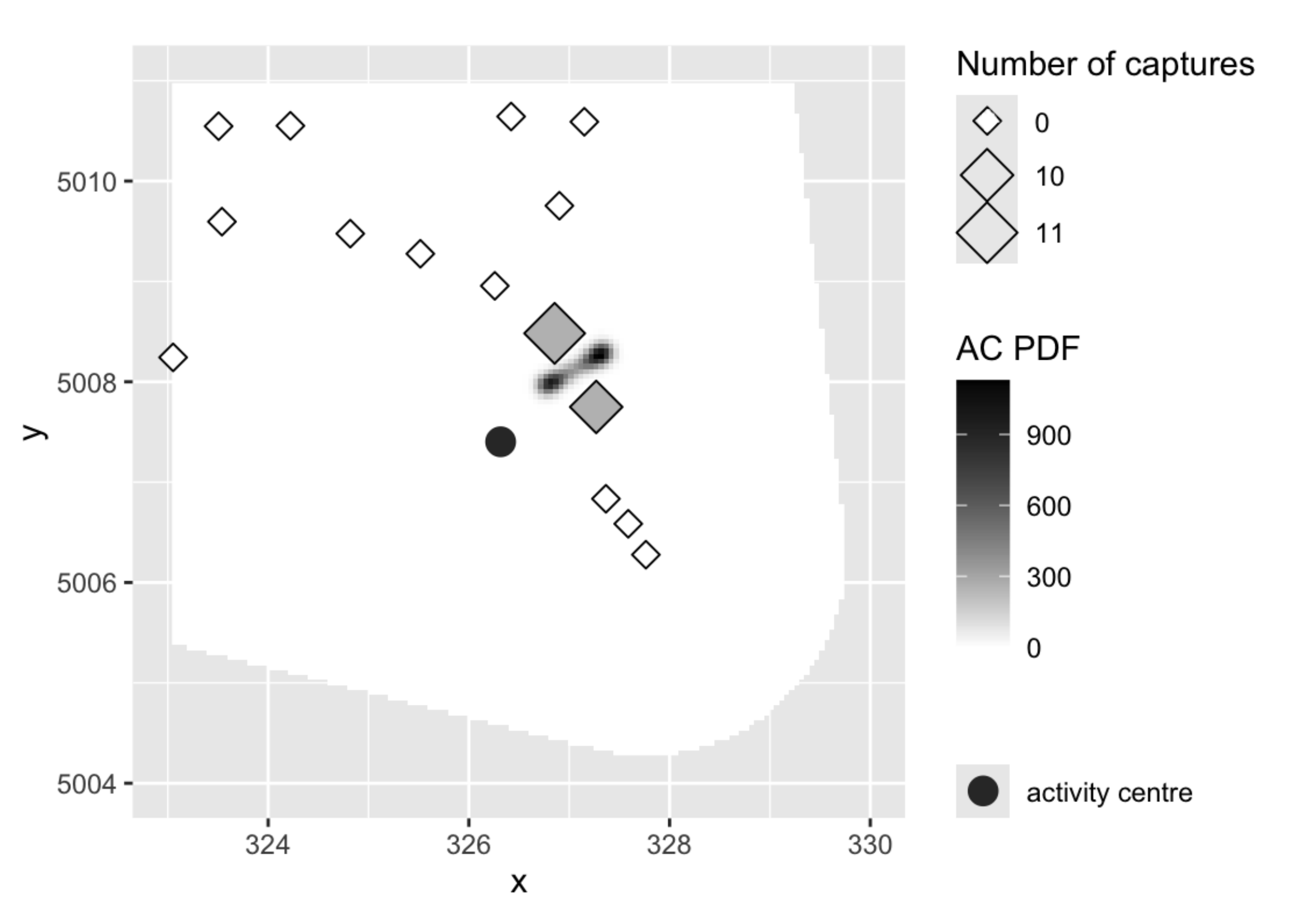}
        \caption{Estimated surface obtained from the CT SCR model for the same simulated individual as in (c).}
        \label{f:sim_surface2_nm}
    \end{subfigure}

    \vspace{10pt}

    \begin{subfigure}[t]{0.495\linewidth}
        \includegraphics[width=\linewidth]{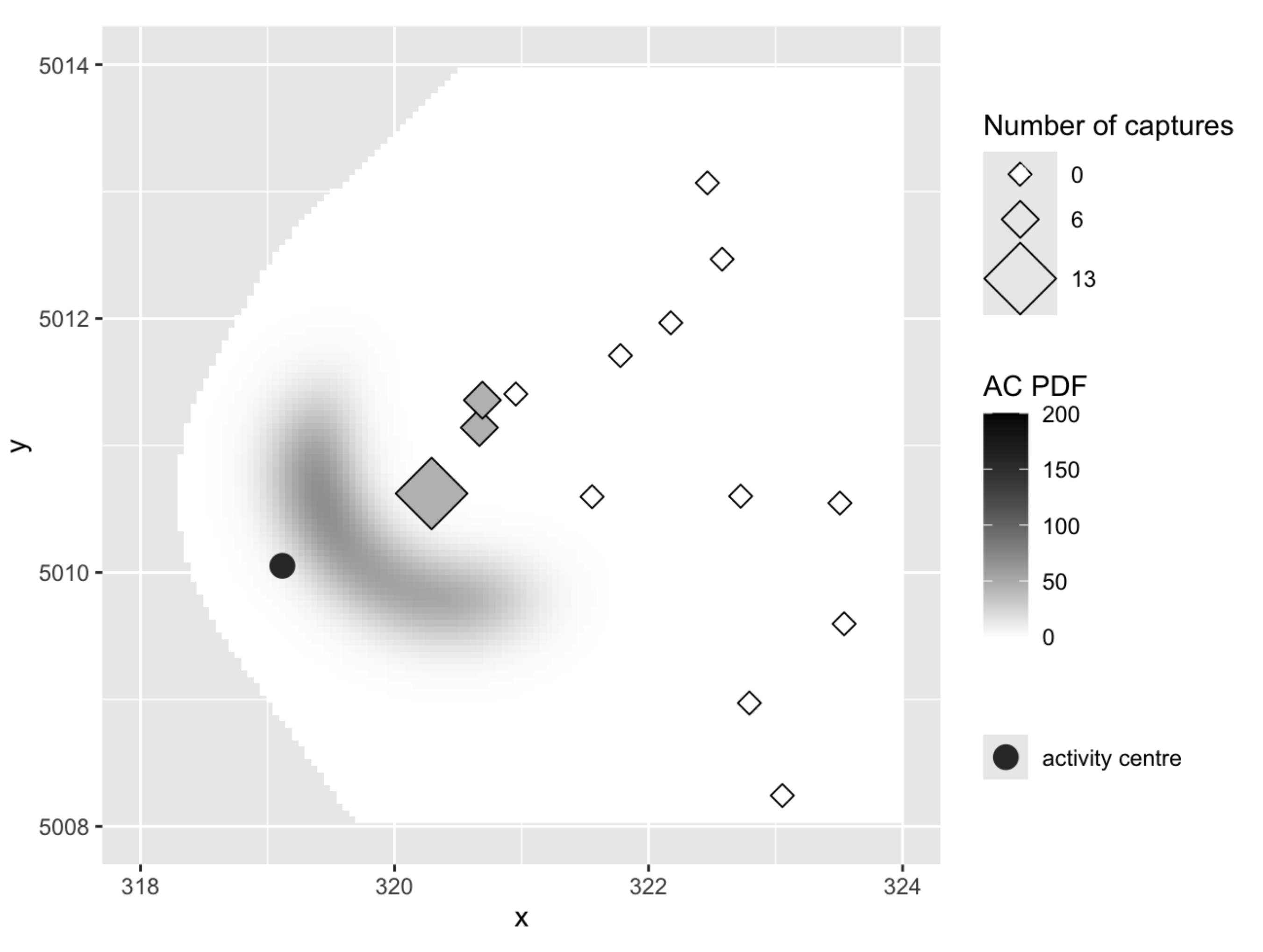}
        \caption{Estimated surface obtained from the MSCR model for a simulated individual.}
        \label{f:sim_surface3}
    \end{subfigure}%
    \hfill
    \begin{subfigure}[t]{0.495\linewidth}
        \includegraphics[width=\linewidth]{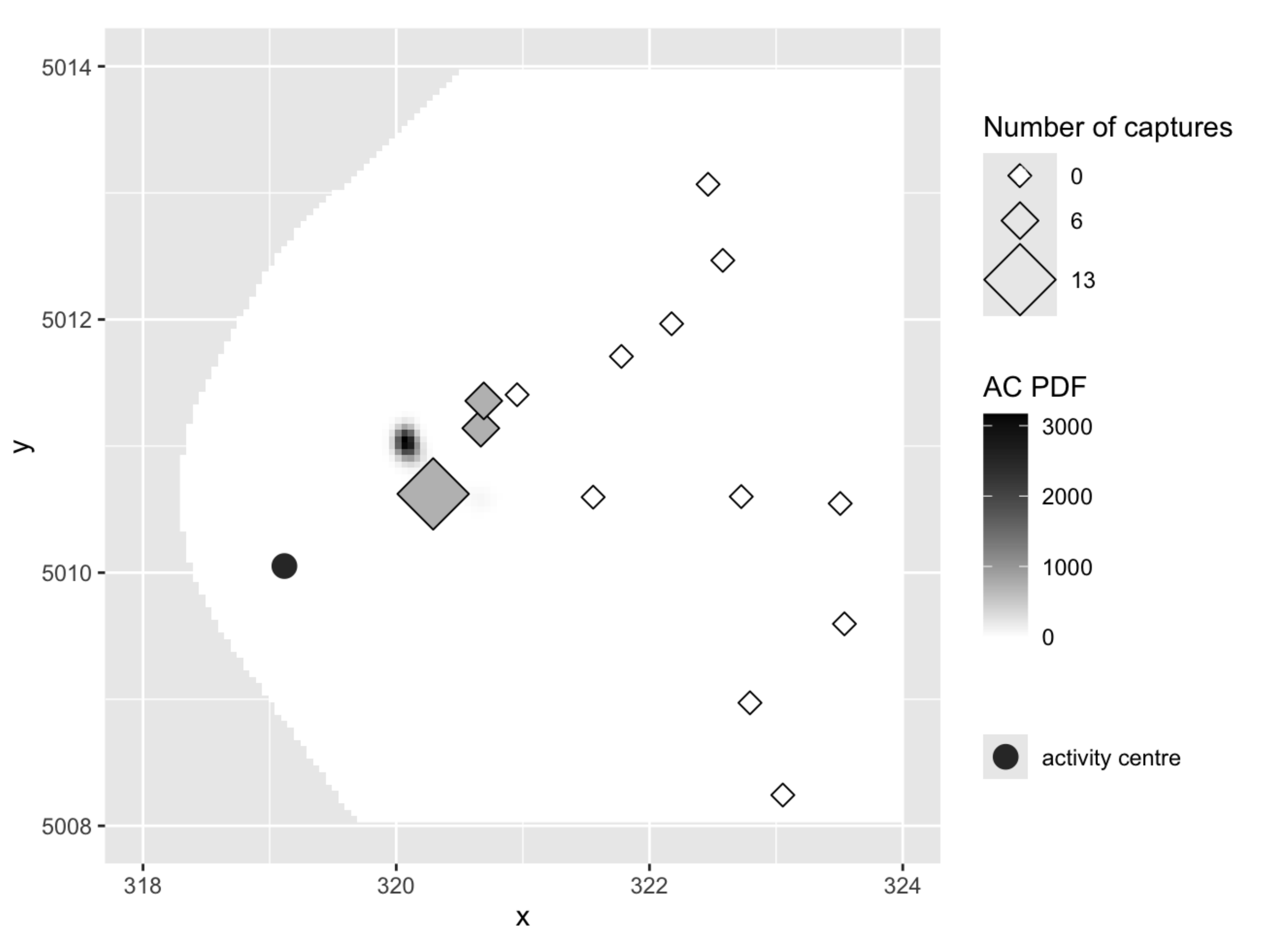}
        \caption{Estimated surface obtained from the CT SCR model for the same simulated individual as in (e).}
        \label{f:sim_surface3_nm}
    \end{subfigure}

    \caption{ Estimated activity centre probability density surfaces from the MSCR ((a), (c), (e)) and SCR ((b), (d), (f)) models for three individuals simulated under the MSCR model. "AC PDF" denotes the activity centre probability density function. In each pair, the MSCR surface assigns positive density near the true activity centre (black point), while the SCR surface typically does not. Map units are in kilometres.}
    \label{f:sim_surfaces}
\end{figure*}

Finally, Figure \ref{f:sim_surfaces} presents the probability densities of the activity centre locations for three simulated individuals. Those individuals were generated as described above, with $\log(\beta) = -1$. Similarly to our case study, the surface densities vary across models for a given individual. We observe that the SCR model produces more concentrated densities, with high values in a small area near the traps where the animal was observed. Those SCR-derived surfaces assign a density of zero around the true activity centre locations (Figures \ref{f:sim_surface1_nm}, \ref{f:sim_surface2_nm}, and \ref{f:sim_surface3_nm}). Meanwhile, the MSCR model results in broader density surfaces that encompass the true simulated activity centre with positive density (Figures \ref{f:sim_surface1}, \ref{f:sim_surface2}, and \ref{f:sim_surface3}). These findings demonstrate the importance of accounting for memory in SCR models, as bias may be introduced when memory effects are ignored.

\section{Discussion}
\label{s:discussion}

Robust population size estimation is essential for conservation, involving both the design and deployment of a data collection protocol combined with an appropriate analysis of the collected data. In this paper, we focused on spatial capture-recapture data collected from camera traps, where exact detection times are recorded. Standard models used to analyze such data consider an individual-specific spatial component, the activity centre, but do not reflect the movement of animals over the course of the survey. To address this critical limitation in SCR methodology, we propose a continuous-time spatial capture-recapture model that explicitly incorporates spatio-temporal correlation between consecutive observations of a single individual.

The proposed model is motivated by the biological reality of animal movement and incorporates the memory of the previous capture of an individual in the probability of a subsequent observation. Memory is incorporated through a hazard function specified as an Ornstein-Uhlenbeck-type process, often used to model animal movement. Following the capture of an individual, the associated future detection risk over space of that individual changes over time, taking into account the known location of the individual at the previous capture as well as its activity centre. Although the common use of the OU process to model animal movement motivated our choice of hazard function, our framework is flexible and allows alternative hazard formulations, such as longer-tailed processes. In the specified model, the standard continuous-time SCR remains a limiting case of MSCR, where the memory component of the model becomes negligible (i.e.~as $\beta \rightarrow \infty$). The fact that standard SCR is a specific case of MSCR is a strong asset of the method, as it places MSCR in the context of recent model extensions and strong theoretical results formerly developed for spatial capture-recapture. We pave the way for a new family of models that considers the fundamental reality of animal movement, without passing over the models that have been developed in recent years.

Our simulation study shows that MSCR consistently provides accurate population size estimates, while standard SCR (without memory) presents substantial negative bias when memory effects are strongly present in the data. As shown in Figure \ref{f:boxplot}, for higher values of $\beta$ both models produce similar accurate results, as we expected from their theoretical equivalence for such parameters. When memory is weak, MSCR and SCR yield similar results. However, only MSCR avoids bias under strong memory scenarios, consistently providing reliable estimates as displayed in Figure \ref{f:boxplot}). Our work presents situations where unmodelled spatio-temporal correlations lead to bias in the parameter estimates, which adds new nuances to previous results such as those of \cite{theng2021confronting}.

The estimated value of $\beta$ for the American marten data set is 0.74 on the log-scale (i.e., $\beta \approx 2.10$). As shown in Figure \ref{f:boxplot}, the simulations indicate that this estimate lies within a range where simulation results show moderate model differences, although not as pronounced as for lower values of $\beta$. More generally, the value of $\beta$ obtained from fitting MSCR to real data sets could be used as an indicator of the amount of spatio-temporal correlation contained within a data set. Understanding the amount of movement captured in a data set is crucial to ensure that the data is treated appropriately and fitted with models that make the correct assumptions. Our findings suggest that data sets that have not been thinned, where individuals are observed multiple times within shorter time intervals over short surveys, will exhibit the most memory effects and require appropriate modelling approaches. We expect the degree of memory present in a data set to be primarily influenced by both the survey design and the species-specific behavior patterns. Furthermore, MSCR eliminates the need to thin data sets before fitting, preserving valuable information and avoiding the arbitrary decisions associated with conventional thinning procedures.

We also demonstrate that models with and without memory result in clear differences in the corresponding activity centre density estimates of the individuals in the population. Accurate activity centre density surfaces are essential for spatial management decisions. The estimates of the probability distributions of activity centres from MSCR were mainly non-overlapping with those of SCR. The latter provided biased estimates for the location of the activity centres in the presence of memory in the generating process. Similar differences in the spatial patterns of the estimated activity centres were observed in the application of the models to the American marten data in Section~\ref{s:application}, despite relatively similar estimates of the total population size (with large overlapping confidence intervals). However, the memory model was a substantially better fit to the data than the non-memory model ($\Delta$AIC = 97.98), suggesting a clearly important memory component intrinsic to the observed data.

By incorporating memory into the SCR framework, MSCR addresses a key ecological limitation of traditional models, their assumption of independence between detections, and opens new directions for spatio-temporal model development. SCR discards the known location of an individual when observed at a given time in the modelling of future observations, and thus ignores the spatial location memory. Our more general modelling approach provides a foundation for further methodological development. For example, the approach can be immediately extended to account for varying camera activity, e.g.~allowing cameras to be inactive for certain periods of the study either due to battery failure or scheduled periods of inactivity, by simply removing the contribution of the given cameras to the all-traps hazard function when they are inactive. Alternatively, the model can also be extended by considering spatial or temporal covariates to model and explain parameter variability (see for example, \cite{distiller2020using}); or considering the population to be open, allowing for population size to change over time due to individuals entering or exiting the population within the study period. For more complex models and/or larger data sets, the computational aspects may need further consideration. In such cases, the numerical integration required over both time and space can have a large computational cost. Other promising avenues include parallelizing the likelihood contributions across individuals or using non-uniform space and/or time discretization, using finer discretization in regions/intervals of high density to reach accurate results through reduced computational effort. Such strategies are the focus of ongoing research and may enhance the scalability of MSCR to larger or more complex ecological systems.

\vspace{1cm}

\section*{Acknowledgements}

We would like to thank Donovan Drummey, Jill Kilborn, and Chris Sutherland for providing access to the American marten spatial capture-recapture data. The collection of this data was supported by funding from the New Hampshire Fish and Game Department and the University of Massachusetts-Amherst.
\\
RK was funded by the Engineering and Physical Sciences Research Council (ESPRC) reference EP/W001616/1. DB and ID were funded by EPSRC reference EP/W002248/1. 
\\
For the purpose of open access, the author has applied a Creative Commons Attribution (CC BY) licence to any Author Accepted Manuscript version arising from this submission.

\vspace{1cm}

\section*{Supplementary Materials}

The American marten data set and all codes necessary for the simulations and the fit of the models are available on GitHub at \newline \url{https://github.com/clarapasu/MSCR}.

\newpage

\appendix

\section{Appendix A}
\label{AppendixA}

We derive the contribution to the likelihood of the capture history of one individual. We consider an individual $i$, its activity centre, $\bfs$, and the last previous capture $\bfw_{i(j-1)} = (t_{i(j-1)},\bfz_{i(j-1)})$. Firstly, we define the all-traps hazard function $h_{\boldsymbol{.}}(t;\bfs,\bfz_{i(j-1)},t_{i(j-1)},\bftheta)$ as the rate of detection over the $K$ traps at time $t$. To obtain it we simply sum the hazards over $k$ as follows
\begin{equation*}
    \label{cumulative_hazard}
    h_{\boldsymbol{.}}(t;\bfs,\bfz_{i(j-1)},t_{i(j-1)},\bftheta) = \sum_{k=1}^K h(\bfz_k,t;\bfs,\bfz_{i(j-1)},t_{i(j-1)},\bftheta).
\end{equation*} 
The probability that the individual remains unobserved from $t_{i(j-1)}$ until the next observation taking place at time $t_{ij}$ is given by the survival function:
\begin{equation}
\label{survival1}
    S(t_{i(j-1)},t_{ij};\bfs,\bfz_{i(j-1)},\bftheta) = \exp \Bigl\{ -\int_{t_{i(j-1)}}^{t_{ij}} h_{\boldsymbol{.}}(t;\bfs,\bfz_{i(j-1)},t_{i(j-1)},\bftheta) dt \Bigl\}.
\end{equation}
The corresponding (recapture) density associated with the next capture $\bfw_{ij}=(t_{ij},\bfz_{ij})$, for $j>1$, is given by
\begin{equation*}
    f(\bfw_{ij}; \bfs, \bfw_{i(j-1)},\bftheta) = S(t_{i(j-1)},t_{ij};\bfs,\bfz_{i(j-1)},\bftheta)h(\bfz_{ij},t_{ij};\bfs,\bfz_{i(j-1)},t_{i(j-1)},\bftheta).
\end{equation*}

For the initial capture of an individual we use the limiting (half normal) hazard function within the derivation of the associated survival function and density function. This assumes that at time $0$ the process is starting at its limiting distribution, given the individuals' activity centres. The density of the first capture, $\bfw_{i1}$, is then given by,
\begin{equation*}
\begin{aligned}
    f(\bfw_{i1}; \bfs,\bftheta) &= \exp \Bigl\{ -\int_{0}^{t_{i1}} \sum_{k=1}^{K} hn(\bfz_k;\bfs,h_0,\sigma^2)dt \Bigl\}hn(\bfz_{i1};\bfs,h_0,\sigma^2)
    \\ &= \exp \Bigl\{ -t_{i1} \ \sum_{k=1}^{K} hn(\bfz_k;\bfs,h_0,\sigma^2)\Bigl\} hn(\bfz_{i1};\bfs,h_0,\sigma^2),
\end{aligned}
\end{equation*}
noting that the study starts at time 0.

The associated contribution to the likelihood for  observed capture history $\bfw_i$ (conditional on the activity centre) is given by,
\begin{equation*}
   f(\bfw_i ; \bfs, \bftheta) = f(\bfw_{i1};\bfs,\bftheta)\Bigl\{ \prod_{j=2}^{J_i} f(\bfw_{ij}; \bfs, \bfw_{i(j-1)},\bftheta) \Bigl\} S(t_{iJ_i},T;\bfs,\bfz_{iJ_i},\bftheta),
\end{equation*}
where the final term denotes the contribution to the likelihood from the final capture time to the end of the study period. Assuming that the activity centres are uniformly distributed over the region $\mathcal{R}$ of area $A$, we can write, 
\begin{equation}
\label{p_capt}
    f(\bfw_i;\bftheta ) = \frac{1}{A} \int_{\mathcal{R}} f(\bfw_i;\bfs,\bftheta)  d\bfs.
\end{equation}

We now consider the probability of being observed at least once during the study, $p(\bftheta)$. Conditional on a known activity center, $\bfs$, and considering the probability of not being observed within the study, which is expressed through the survival function, we write
 
\begin{equation*}
\begin{aligned}
    1 - p(\bftheta ; \bfs) &=  \exp \Bigl\{ -\int_{0}^{T}hn_{\boldsymbol{.}}(\bfs,h_0,\sigma^2)dt \Bigl\} \\
    \Rightarrow \qquad p(\bftheta ; \bfs) &= 1- \exp\Bigl\{-T \ hn_{\boldsymbol{.}}(\bfs,h_0,\sigma^2)\Bigl\}.
    \end{aligned}
\end{equation*}
We note that once again we assume the limiting distribution for the hazard function within this formulation (as there is no previous known location). 
Once more integrating out over all possible activity centers we obtain,
\begin{equation}
\label{p_obs}
    p(\bftheta) = \frac{1}{A} \int_{\mathcal{R}} p(\bftheta ; \bfs) d\bfs.
\end{equation}

The likelihood terms in Equation \ref{p_capt} and \ref{p_obs} can only be expressed as a function of (multiple) analytically intractable integrals over space and time.

\label{lastpage}

\end{document}